# Shallow-flow velocity predictions using discontinuous Galerkin solutions

Georges Kesserwani, Ph.D[1]; Janice Lynn Ayog[2]; Mohammad Kazem Sharifian Ph.D[3]; and Domenico Baú, Ph.D[4]

[1]Research Fellow and Senior Lecturer, Department of Civil and Structural Engineering, University of Sheffield, Mappin St, Sheffield City Centre, Sheffield S1 3JD, UK. Email: g.kesserwani@sheffield.ac.uk (corresponding author)

[2]Ph.D Candidate, Department of Civil and Structural Engineering, University of Sheffield, Mappin St, Sheffield City Centre, Sheffield S1 3JD, UK. Email: jay@ums.edu.my

[4]Research Associate, Department of Civil and Structural Engineering, University of Sheffield, Mappin St, Sheffield City Centre, Sheffield S1 3JD, UK. Email: g.kesserwani@sheffield.ac.uk

[4]Senior Lecturer, Department of Civil and Structural Engineering, University of Sheffield, Mappin St, Sheffield City Centre, Sheffield S1 3JD, UK. Email: d.bau@sheffield.ac.uk

**Abstract**
Numerical solvers of the two-dimensional (2D) shallow water equations (2D-SWE) can be an efficient option to predict spatial distribution of velocity fields in quasi-steady flows past or throughout hydraulic engineering structures. A second-order finite volume solver (FV2) spuriously elongates small-scale recirculating eddies within its predictions, unless sustained by an artificial eddy viscosity, while a third-order finite volume (FV3) solver can distort the eddies within its predictions. The extra complexity in a second-order discontinuous Galerkin (DG2) solver leads to significantly reduced error dissipation and improved predictions at a coarser resolution, making it a viable contender to acquire velocity predictions in shallow flows. This paper analyses this predictive capability for a grid-based, open source DG2 solver with reference to FV2 or FV3 solvers for simulating velocity magnitude and direction at the sub-meter scale. The simulated predictions are assessed against measured velocity data for four experimental test cases. The results consistently indicate that the DG2 solver is a competitive choice to efficiently produce more accurate velocity distributions for the simulations dominated by smooth flow regions.

**Practical Applications**

The estimation of the spatial distribution of horizontal velocity fields is useful to analyse the design of hydraulic and flood-defence structures undergoing shallow water flow processes. Examples include flooding through a residential area with piered buildings where recirculation flow eddies occur within side cavities, past building blocks, and across of street junctions. This paper demonstrates the utility of a relatively new hydraulic simulation tool, so-called DG2 solver, as an alternative to existing finite volume solvers featured in popular tools such as HEC-RAS 2D, Rubar20, Iber and TUFLOW-HPC. The DG2 solver is open source, as part of the LISFLOOD-FP 8.0 software suite, and particularly excels in the estimation of velocity fields that are more informative of the flow processes within a few minutes of simulation time. The proposed simulation tool is limited to modelling problems where the depth-integrated assumption of the shallow water equations is appropriate.

**Introduction**

Detailed spatial distribution of steady velocity fields is useful to support hydraulic engineering designs and applications, such as transport analyses for predicting the accumulation of woody debris around bridge piers (Mazur et al. 2016; Pagliara and Carnacina 2013; Ruiz-Villanueva et al. 2017) and the distribution of suspended sediments and solute particles in large river system (Dinehart and Burau 2005; Jodeau et al. 2008; Legleiter and Kinzel 2020). To obtain detailed spatial velocity fields, direct in-situ measurements are not often feasible due to safety concerns and lack of access to surveying equipment. Instead, the velocity fields are analysed in physical models that are designed to replicate real-world hydrodynamics, which is useful when analysing complex wave features around and throughout specific engineering structures. For instance, physical models have been used to study the recirculation patterns of velocity fields in lateral channel cavities (Jackson et al. 2015; Juez et al. 2018; Mignot and Brevis 2020), diverting junctions (Mignot et al. 2013; Momplot et al. 2017; Shettar and Murthy 1996) and around dense piered-building blocks in urban residential areas (Chen et al. 2018; LaRocque et al. 2013; Smith et al. 2016). As complex physical models tend to be costly, computational models can be used as a direct alternative to them, or to support the analysis of data from physical laboratory experiments.

To avoid the prohibitive cost of three-dimensional (3D) Navier-Stokes computational models, numerical solvers of the basic form of the 2D depth-integrated shallow water equations (SWE) can be used to efficiently produce sufficiently detailed velocity fields at sub-meter grid resolution – at least for hydraulic engineering applications where the vertical water depth is very shallow, and the variation of the velocity is mostly uniform in the vertical direction (Duan 2005; Aureli et al. 2015). In this context, 2D-SWE solvers introduce a simplified turbulence model via the Manning's friction formula (Bonetti et al., 2017; Gioia and Bombardelli, 2002), and are often built around second-order finite volume (FV2) numerical methods (Wang et al. 2013), of which Flood Modeller (JACOBS 2022), TELEMAC (Galland et al. 1991), HEC-RAS 2D (USACE 2021), Rubar20 (INRAE 2021), Iber (Bladé et al. 2014) and TUFLOW-HPC (BMT-WBM 2018). Even on a fine grid resolution, FV2 solvers tend to develop a fast growth rate of dissipation error that somewhat allows to approximate turbulence effects in the complex regions of flow-structure interaction.

However, the magnitude of the dissipation error with FV2 solvers can be too large for it to also be able to capture the flow phenomena occurring in smooth shallow-flow regions such as recirculation eddies within lateral channel cavities (Kimura and Hosoda 2007; Ouro et al. 2020; Ozgen-Xian et al. 2021), past obstacles due to vortex shedding (Kim et al. 2009), and past piers in bridges and buildings (Horritt et al. 2006). For example, FV2 solver predictions can overly elongate or even smear out small-scale recirculation eddies (Cea et al. 2007; Navas-Montilla et al. 2019). This adverse effect can be alleviated by adding artificial eddy viscosity to the 2D-SWE (Collecutt and Syme 2017; Syme 2008), but this approach trades off with case-dependent and often quite onerous calibration efforts (Bazin 2013; Navas-Montilla et al. 2019). Another alternative is to use third-order finite volume (FV3) solvers, which have a reduced rate of dissipation error (Navas-Montilla et al. 2019), but FV3 solvers can still lead to distorted eddy predictions such as in the far wake past an obstacle as shown in Macías et al. (2020).

A Discontinuous Galerkin (DG) numerical solver is numerically more complex compared to an equally accurate finite volume (FV) solver leading to: (i) improved velocity predictions for shallow flow at a coarser resolution (Kesserwani 2013; Kesserwani and Wang 2014), and (ii) significantly smaller error dissipation at a fixed resolution (Zhou et al. 2001; Aizinger and Dawson 2002; Zhang and Shu 2005; Wang and Liu 2005; van den Abeele et al. 2007; Duran and Marche 2014; Schaal et al. 2015; Ayog et al. 2021). These imply that a DG solver of the 2D-SWE could inherently provide better quality capture of the phenomena occurring in the smooth flow regions and be tuned, via resolution coarsening, to retain sufficient error dissipation to approximate turbulence effects in the complex regions of flow-structure interaction. This aspect is investigated for a robust, grid-based second-order DG (DG2) 2D-SWE solver that is parallelised on Graphical Processing Units (GPU) and openly accessible on LISFLOOD-FP 8.0 (Shaw et al. 2021).

For realistic hydraulic simulations, DG2 solvers of the 2D-SWE have often been formulated to keep a sensible balance between numerical complexity, computational cost, robust integration of steep bed slope and friction terms with wetting and drying, and GPU memory affordability (Lambrechts et al. 2010; Le Bars et al. 2016; Clare et al. 2021; Le et al. 2020; Kärnä et al. 2018; Kesserwani et al. 2018; Mulamba et al. 2019; Pham Van et al. 2016; Winters and Ghassner 2015; Wintermeyer et al., 2018; Wood et al. 2020; Kesserwani and Sharifian 2020; Shaw et al. 2021; Wu et al. 2021). Ayog et al. (2021) benchmarked the DG2 solver against FV industrial solvers for realistic flood modelling. The authors found that the DG2 solver can reproduce the water level hydrographs of the FV-based solvers at twice to four times coarser resolution, with similar findings reported in Shaw et al. (2021) for a catchment-scale flood simulation. Compared to the FV and other solvers, the DG2 solver could detect the presence of small-scale velocity variations with unrivalled accuracy (Ayog et al. 2021; Shaw et al. 2021). Few other studies have also hinted at the potential of DG2 solvers to capturing small-scale flow phenomena in the velocity predictions. Kubatko et al. (2006) studied DG2 and higher-order DG solvers in reproducing tidal flow fields over an idealised channel, concluding that the DG2 solver delivers well-captured eddies as accurately as the higher-order DG solvers. Alvarez-Vázquez et al. (2008) applied a DG2 solver to simulate fish migration in coastal flows, also commenting on its ability to capture eddies. Beisiegel et al. (2020) applied a DG2 solver to simulate hurricane-induced flow circulation in coastal areas, showing that it can replicate the asymmetrical patterns of the recirculation eddies extracted from a 3D Navier-Stokes computational model. Still, a dedicated study is needed to understand the extent to which a DG2 solver can reliably predict shallow-flow velocity fields at sub-meter grid resolution, which are featured by quasi-steady flow phenomena in and around hydraulic structures.

This paper studies this aspect for the grid-based DG2 solver of the 2D-SWE (Shaw et al. 2021) in relation to reproducing spatially horizontal velocity fields for quasi-steady experimental test cases. In the next section, the DG2 solver and equally accurate FV2 solver (Ayog et al. 2021) are overviewed as well as the selected test cases that their set-up files are made openly available (Ayog and Sharifian 2022). The test cases are used to investigate flow phenomena with vortex shedding past a submerged conical island, recirculation zones in lateral cavities, flow diversion into a junction including the presence of an obstacle in the main branch, and a flood flow across and around piered buildings. The DG2 predicted velocity fields are compared to those predicted by FV2 or FV3 solvers and against experimentally measured velocities. The comparison includes qualitative and quantitative analysis to assess the reliability in the solver predictions and the impact of resolution coarsening at the sub-meter scale. Conclusions are finally drawn to recommend when it is best to use the DG2 solver to produce reliable velocity predictions within the scope of depth-integrated 2D-SWE modelling.

**Materials and Methods**

Finite difference, finite element and FV numerical methods – including Alternating Direction Implicit (ADI) solvers (JACOBS 2022)– have been developed to solve hyperbolic partial differential equations casting the mass and momentum conservation laws of the 2D-SWE (Galland et al. 1991; Bassi and Rebay 1997; Cockburn and Shu 2001; Alcrudo and García-Navarro 1993; Guinot 2003; Toro 2001; Zhou et al. 2001; Toro and García-Navarro 2007).

**Table 1** Differences in the level of numerical complexity and implementation features between the DG2 and FV2 solvers of the depth-integrated 2D-SWE.

| Flow solver | DG2 (Shaw et al., 2021) | FV2 (Ayog et al., 2021) |
|---|---|---|
| Mathematical model | Partial differential equations governing shallow water flows<br>$$\partial_t \mathbf{U} + \partial_x \mathbf{F}(\mathbf{U}) + \partial_y \mathbf{G}(\mathbf{U}) = \mathbf{S_b}(\mathbf{U}) + \mathbf{S_f}(\mathbf{U})$$<br>$\partial$ represents a partial derivative operator<br>$\mathbf{U}(x, y, t) = [h, q_x, q_y]^T$ is the vector of flow variables<br>$\mathbf{F} = [q_x, q_x^2/h + gh^2/2, q_x q_y/h]^T$ and $\mathbf{G} = [q_y, q_x q_y/h, q_y^2/h + gh^2/2]^T$ are vectors representing the components of physical flux field<br>$\mathbf{S_b} = [0, -gh\partial_x z, -gh\partial_y z]^T$ and $\mathbf{S_f} = [0, -C_f u\sqrt{u^2+v^2}, -C_f v\sqrt{u^2+v^2}]^T$ are the bed and friction source term vectors<br><br>$t$ is time (s) and $g$ is the gravity acceleration (m$^2$/s)<br>$(x, y)$ the horizontal position and $z$ the bed elevation (in m)<br>$h$ is the water depth (m)<br>$(q_x, q_y) = (hu, hv)$ are unit-width flow discharges (m$^2$/s)<br>$(u, v)$ are depth-averaged components of the velocity field (m/s)<br>$C_f = g n_M^2 / h^{1/3}$ is a function of the Manning's coefficient $n_M$ (s m$^{-1/3}$) | |
| Shape of $\mathbf{U}_h$ over a grid element | Planar variation that is spanned by an average coefficient, $\mathbf{U}_c^0$, and two slope coefficients, $\mathbf{U}_c^{1x}$ and $\mathbf{U}_c^{1y}$. $\mathbf{U}_h(x,y,t)$<br>$= \mathbf{U}_c^0(t) + \frac{2\sqrt{3}(x-x_c)}{\Delta x}\mathbf{U}_c^{1x}(t)$<br>$+ \frac{2\sqrt{3}(y-y_c)}{\Delta y}\mathbf{U}_c^{1y}(t)$ | Constant variation spanned by one average coefficient $\mathbf{U}_c^0$ |
| Slope limiting | Local limiting to the intrinsic slope coefficients after spatial update | Global limiting to extrinsically reconstructed slopes from the piecewise averaged flow data |
| HLL Riemann flux calculations | Called 6 times, 2 per coefficient | Called only 2 times |
| Terrain integration with wet-dry front treatments | Pre-processing the digital elevation model into piecewise planar projection, with three coefficients, using a toolkit (Shaw et al. 2021). Well-balanced reconstruction for the average and the slope coefficients ensuring water depth positivity (Kesserwani and Liang 2012) | Planar constant representation of the digital elevation model. Well-balanced reconstructions at the faces ensuring water depth positivity (Liang and Marche 2009) |
| Friction integration | Split implicit scheme to update the average and slope coefficients of the discharges | Split implicit scheme to update averaged discharges |
| Two stages Runge-Kutta integration | Courant number = 0.3 (Cockburn and Shu 2001) | Courant number = 0.5 (Kesserwani and Liang 2012) |
| Software environment | Both codes can be run from LISFLOOD-FP 8.0 using the local repository of the University of Sheffield (2021). | |
| Initial and boundary conditions | Set-up in line with the general guidance for running LISFLOOD-FP 8.0 (www.seamlesswave.com/LISFLOOD8.0). | |

These include FV and discontinuous Galerkin methods using the Godunov (1959) approach introduced in aerodynamics and gas dynamics (Bassi and Rebay 1997; van Leer 2006; Toro 1989; Bernetti et al. 2008). For realistic hydraulic simulations, second-order accurate formulations have often been adopted to incorporate robustness features (Toro and García-Navarro 2007; García-Navarro et al. 2019; Winters and Ghassner 2015; Kesserwani et al. 2018), as is the case with the FV2 and DG2 solvers investigated in this work (Ayog et al. 2021).

Both, the FV2 and DG2, solvers update flow data (i.e., the water depth and the components of the horizontal unit-width discharge) elementwise from the inter-elemental spatial flux exchange of the Riemann problem solutions (Toro & Garcia-Navarro, 2007). In contrast to the FV2 solver, which stores and evolves one coefficient of a piecewise-averaged flow data, the DG2 solver does the same for three coefficients – of an average and two directionally independent slopes – defining piecewise-planar flow data. The elementwise storage and evolution of the DG2 solver's slope coefficients adds numerical complexity, which can be further simplified to conveniently make the DG2 solver operate on a similar calculation stencil to that of the FV2 solver (Kesserwani et al., 2018). Table 1 summarises the difference in the flow data structure, the numerical complexity and the implementation treatments featuring in the FV2 and DG2 solvers of the 2D-SWE.

### *Test cases and quantitative indices*

Test cases with experimental velocity fields are selected for hydraulic modelling problems characterised by shallow water flows through and past structures. The properties of the test cases are summarised in Table 2. The initial and boundary conditions and digital elevation models to set up the simulations for the test cases are openly available (Ayog and Sharifian 2022). For the second to the fourth tests, the numerically predicted velocity vectors and cross-sectional profiles are analysed with respect to experimentally measured velocity data, quantitatively via the $R^2$ coefficient and the $L^1$-norm error. These indices are expressed as:

$$R^2 = \left[ \frac{\sum_{k=1}^{N_s} (V_k^{EXP} - \underline{V}^{EXP})(V_k^{NUM} - \underline{V}^{NUM})}{\sqrt{\sum_{k=1}^{N_s} (V_k^{EXP} - \underline{V}^{EXP})^2 \sum_{k=1}^{N_s} (V_k^{NUM} - \underline{V}^{NUM})^2}} \right] \quad (1)$$

$$L^1 - norm\ error = \frac{1}{N_s} \left( \sum_{k=1}^{N_s} |V_k^{EXP} - V_k^{NUM}| \right) \quad (2)$$

In Eqs. (1-2), $V^{EXP}$ and $V^{NUM}$ are the velocity magnitudes from the experimental measurements and numerical predictions, whereas $\underline{V}^{EXP}$ and $\underline{V}^{NUM}$ represent their space-averaged values. $N_s$ denotes the total number of sampled data ($k = 1, … N_s$). The indices for the cross-sectional spatial profiles are calculated in a similar manner to the velocity vectors, but by using the spatial point values instead of the velocity magnitudes. The $R^2$ coefficient can vary between 0 and 1, and quantifies the statistical correlation, or similarity, between the numerically predicted profiles, or velocity magnitudes, and those from the experiments. The $L^1$-norm error provides an estimate of the discrepancy between the simulated values and the experimentally measured data. Besides these indices, the relevance index (RI) is used to analyse the directional alignment between the predicted and measured velocity fields, taking a value of 1 with perfect alignment and 0 otherwise. It is expressed as:

$$RI = \frac{1}{N_s}\left(\sum_{k=1}^{N_s} \frac{\left(u_k^{EXP} \times u_k^{NUM}\right) + \left(v_k^{EXP} \times v_k^{NUM}\right)}{V_k^{EXP} V_k^{NUM}}\right) \qquad (3)$$

where $u^{EXP}$ and $v^{EXP}$ are the velocity components from the measured velocities and $u^{NUM}$ and $v^{NUM}$ are those from the numerical predictions.

**Table 2.** Selected test cases with experimentally measured velocity data. For each test case, the vertical water depth *h* (m) and smallest horizontal length *L* (m) are listed to obtain the aspect ratio *r* for the validation of the depth-integrated shallow water equations.

| Test case | Velocity data type | h | L | r |
|---|---|---|---|---|
| Vortex shedding past a conical island | Instantaneous horizontal velocity fields and time series for the velocity components at two gauge points (Lloyd and Stansby 1997). http://coastal.usc.edu/currents_workshop/problems/prob1 | 0.054 | 1.52 | 0.036 |
| Recirculation flow in sharp building cavities | Time-averaged spatial velocity fields from particle image velocimetry (Rubinato et al. 2021). | 0.023 | 1.18 | 0.019 |
| Flow diversion at a T-junction | Measured recirculation flow width of the time-averaged velocity fields, and cross-sections of velocity components at the junction (Bazin 2013; Bazin et al. 2017). | 0.046 | 0.3 | 0.153 |
| Flooding in an urban residential area | Maximum horizontal velocity vectors from vane probes and velocity meter (Smith et al. 2016). http://datawarehouse.wrl.unsw.edu.au/newcastlefloodmodel | 0.5 | 120 | 0.004 |

**Results and Discussion**

*Vortex shedding past a conical island*

This test involves vortical structures arising from a steady, subcritical inflow past a submerged conical island. It has a snapshot of measured 2D instantaneous vector fields, produced by particle tracking velocimetry (Lloyd and Stansby 1997), and time series of the horizontal velocity components at two gauging points past the island, point 1 (6.02 m, 0.76 m) and point 2 (6.02 m, 1.03 m). The test setup, as shown in Fig. 1, consists of a channel that is 9.75 m long and 1.52 m wide with a flat bed, on which a conical island is placed 5 m downstream from the inflow boundary. The island has a height of 0.049 m, a side slope of 8.0 degrees, and a diameter of 0.05 and 0.75 m at the top and base of the conical island, respectively. A unidirectional inflow velocity of $u = 0.115$ m/s inundates the island until it becomes fully submerged under a water depth of 0.054 m. The DG2 solver was run until 500 s at the grid resolution of 0.0152 m (medium), specified by Lloyd and Stansby (1997), and considering twice finer (fine) and coarser (coarse) resolutions of 0.0076 and 0.0304 m, respectively. The Manning's coefficient was calibrated, considering optimal tracking of the measured velocity component signals at the gauge points, to $n_M = 0.014$ s m$^{-1/3}$ that is similar to the value obtained in alternative studies (NTHMP 2016; Lynett et al. 2017).

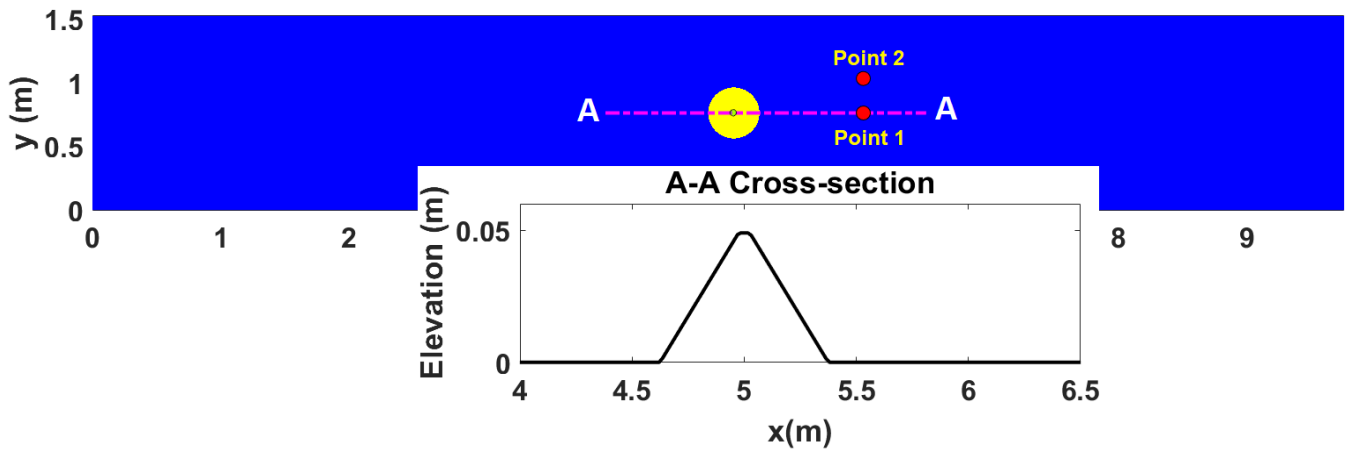

Fig. 1. Layout of the conical island experiment made of a flat topography (blue area) and a cone-shaped island with a 0.75 m diameter base surface (yellow circle) and a 0.05 m diameter top surface (small green circle). A cross-section of the conical island is extracted across the centreline A-A and shown in the insert plot below. Two gauging points (points 1 and 2) record the time histories of the velocity components.

Existing studies already uncovered the weakness of the FV2 solver to capture vortex formation past a cylindrical obstacle with reference to depth-integrated solvers including more complex turbulence models (e.g., Kim et al. 2009; NTHMP 2016; Ginting and Ginting 2019). The FV2 solver predictions lead to significantly elongated recirculation length with slower velocity recovery rate past the obstacle, but these shortcomings can be improved with the DG2 solver predictions (Sun et al. 2022). Moreover, the DG2 solver predictions seem to be more reliable than those carried out by an FV3 solver for this test case (Macías et al. 2020). As shown in Fig. 2, the FV3 solver leads to spurious eddy asymmetricity in the far wake region past the cylindrical island that does not occur with the DG2 predicted velocity fields. Therefore, only the DG2 predicted velocity fields are investigated next compared to the measured fields and *v*-velocity time series for the fine, medium and coarse grid resolutions.

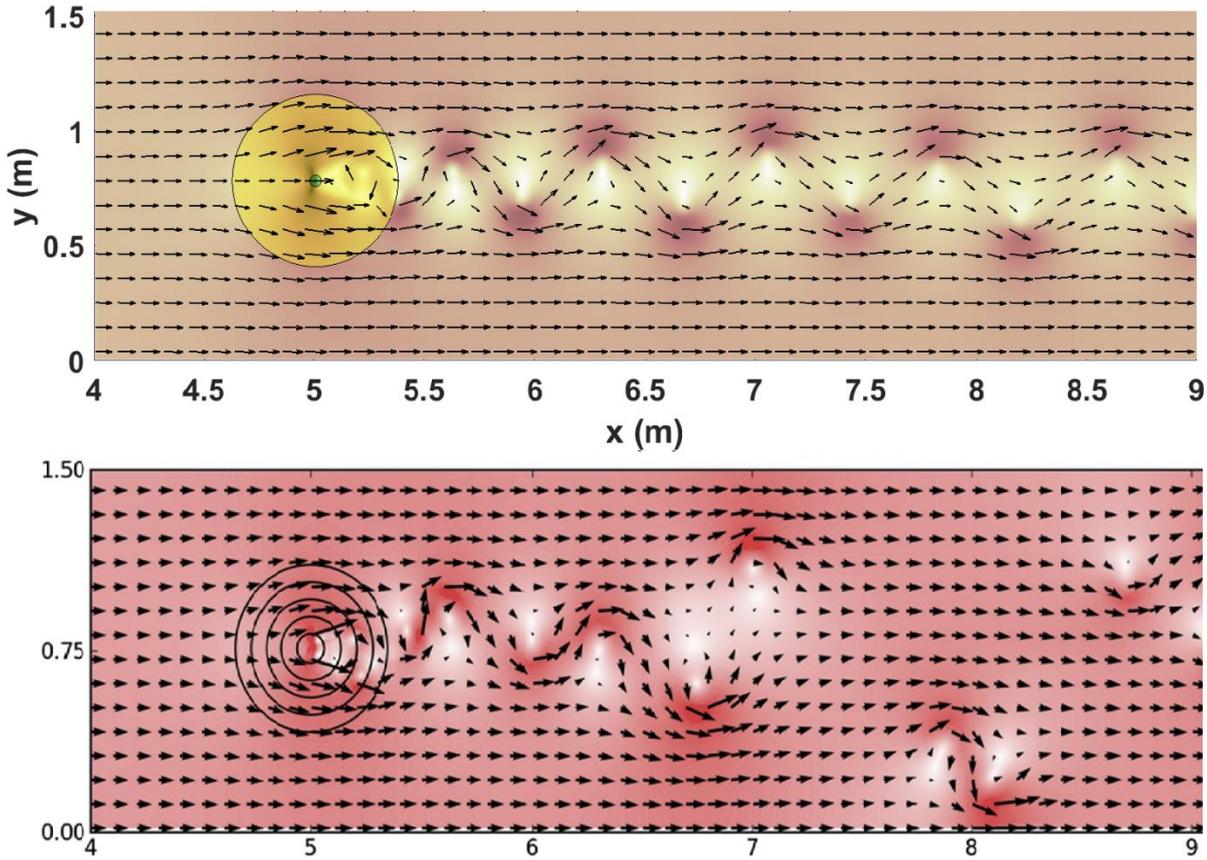

Fig. 2. Velocity vectors produced by the DG2 (top panel) and the FV3 solver in Macías et al. (2020) (bottom panel). The FV3 predicted velocity vectors exhibit short vortex street, with spurious vortices detaching and spreading asymmetrically in the far wake region. In contrast, those predicted by the DG2 solver are consistent with the patterns predicted by a more complex 2D depth-averaged Reynolds-averaged Navier–Stokes model (Ginting and Ginting, 2019).

Fig. 3 includes the instantaneous velocity vectors extracted from the velocity fields predicted by the DG2 to the measured vectors (Fig. 3a). Following Kim et al. (2009), the wake region from $x = 5$ m (centre of the island) to $x = 6.2$ m (downstream of the gauging points) is equally divided into three subregions, each with $x$-directional length of 0.375 m corresponding to the radius of the island's base. In subregion 1, the shaded vicinity past the conical island, the DG2 solver at the coarse resolution (Fig. 3d) predicts the best directional alignment with the measured vector fields (Fig. 3a) where it shows a single eddy. This eddy becomes more defined at the medium resolution (Fig. 3c) but gets distorted into two eddies at the fine resolution (Fig. 3b). Similar is observed for subregion 2, where the DG2 solver again predicts the best directional alignment at the coarse and medium resolutions (Fig. 3c-d) but generates a spurious shift in the directions for the fine resolution (Fig. 3b). In subregion 3, the DG2 solver correctly predicts a unidirectional flow in the vicinity of point 2 for all the resolutions (Fig. 3b-d). However, at point 1, its predicted directional alignments better match the directions seen in the measured vectors at the coarse resolution (Fig. 3d). At the medium resolution (Fig. 3c), the predicted fields are seen to be crossing point 1 in the wrong direction arising from the presence of a spurious eddy. At the fine resolution (Fig. 3b), the solver also predicts a spurious eddy with an upward shift in its location, leading to better directionality alignment with the measured vectors around point 1.

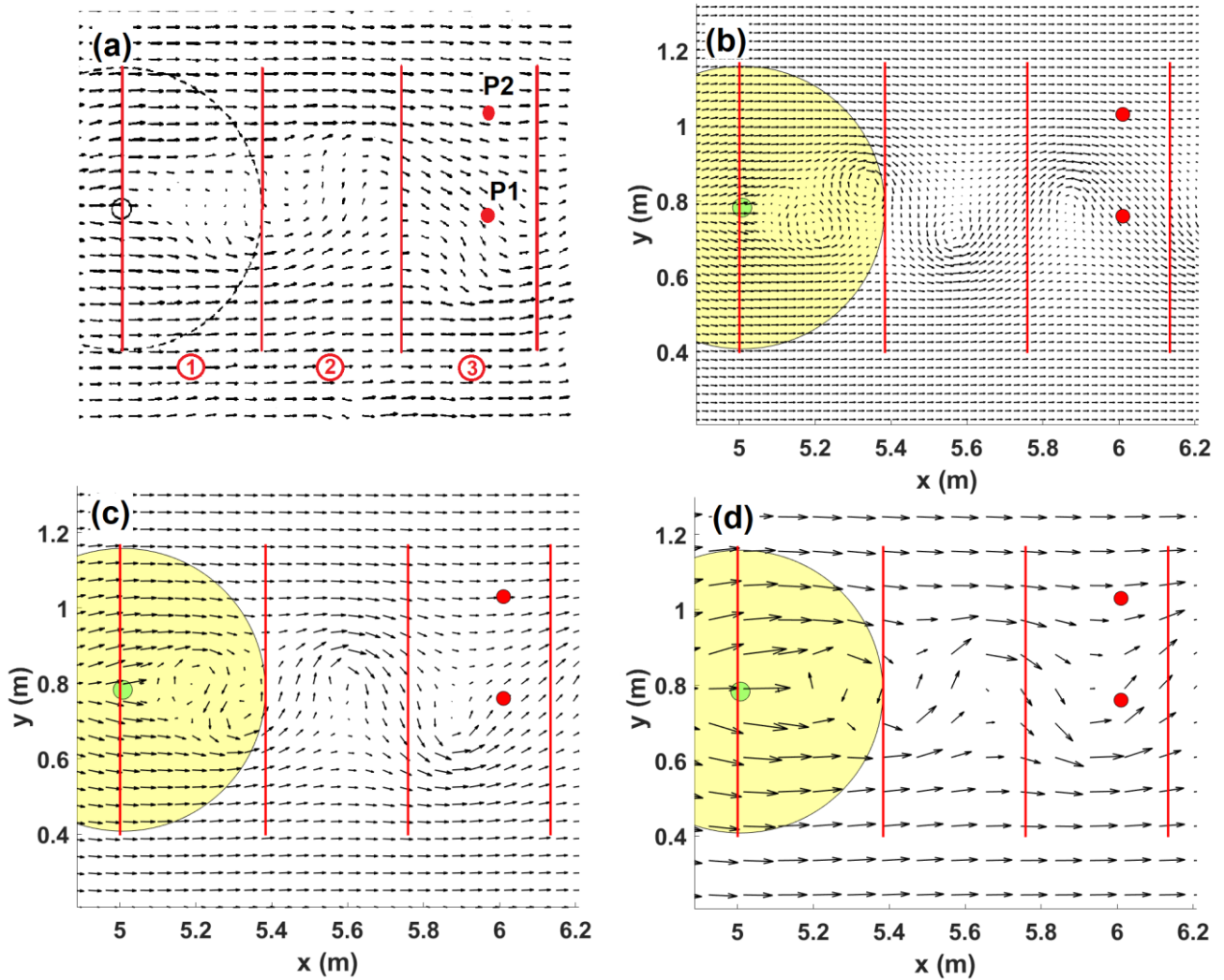

Fig. 3. Velocity vector fields from: (a) the experimental measurements of Lloyd and Stansby (1997), and from the DG2 solver predictions at (b) the fine resolution, (c) the medium resolution and (d) the coarse resolution. The red dots indicate the location of the gauging points, and the four red lines divides the wake into three subregions (red circled numbers), each with a distance equal to the radius of the island's base surface (0.375 m).

The identified shortcoming of the DG2 solver predictions at the medium and fine resolutions can also be seen in the extracted time series of the $v$ component of the velocity at the gauge points. As shown in Fig. 4, the DG2 solver at the coarse resolution provides close enough match to the measured time series but induces larger amplitudes for the medium and, more noticeably, the fine resolution. It can therefore be inferred that the DG2 solver tends to more reliably approximate complex vortex shedding structures for the coarse resolution but would need a turbulence model as the grid resolution is refined. This is particularly encouraging in terms of runtime cost as the solver needed half a minute to complete the simulation for the coarse resolution, whereas it took around 4 and 31 minutes for the medium and fine resolution, respectively.

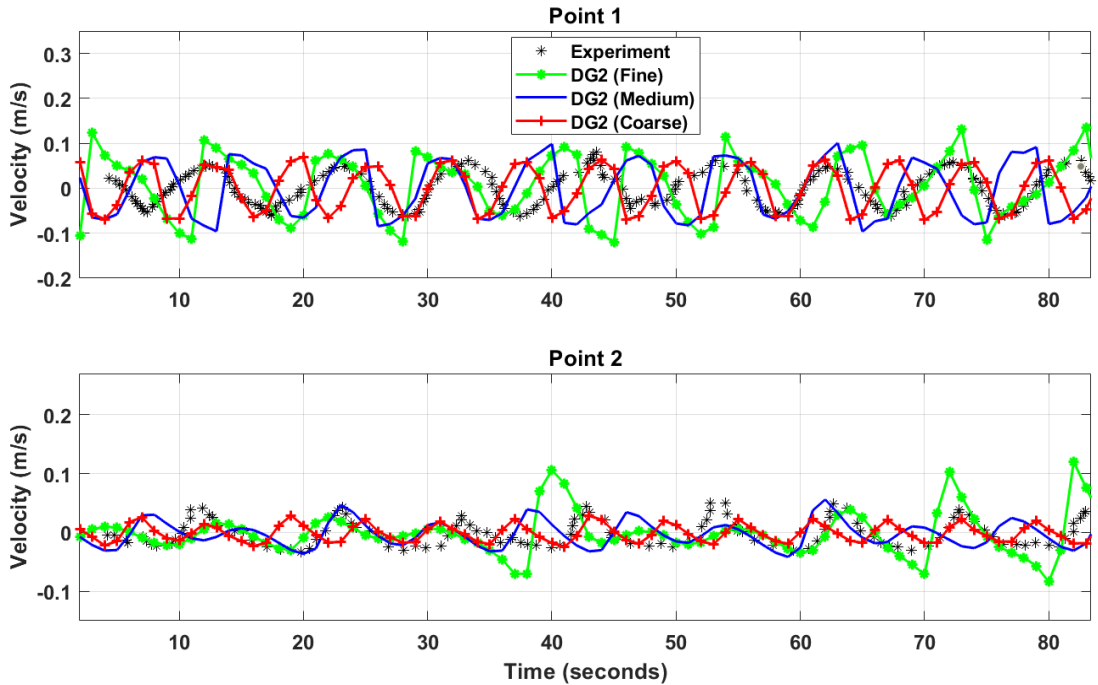

Fig. 4. Time histories of the DG2 predicted $v$ component of the velocity at the different resolution compared to the experimental measurement at the two gauging points.

### *Recirculation flow in sharp building cavities*

The ability of the DG2 and FV2 solvers is examined to reproduce time-averaged spatial velocity fields from high-resolution experimental data obtained from particle image velocimetry (PIV) measurements performed in a physical model in the University of Sheffield (Rubinato et al. 2021). The test case configuration is installed on top of an existing experimental flume and is illustrated in Fig. 5a (Rubinato 2015). The flume slopes at 0.001 m/m and has a smooth surface ($n_M = 0.011$ s m$^{-1/3}$), where a quasi-steady flow develops. The flow is driven by a skewed inflow, extracted from the measured velocity fields at flume's southern boundary (red box, Fig. 5a). At the northern boundary of the flume, free outflow numerical conditions are imposed. The DG2 and FV2 solvers were run at a resolution of 0.016 m (fine), matching the resolution of the PIV data, and at a twice coarser resolution of 0.032 m (medium). Simulation results at a four times coarser resolution of 0.064 m (coarse) led to uncompetitive predictions and will not therefore be investigated next (not shown).

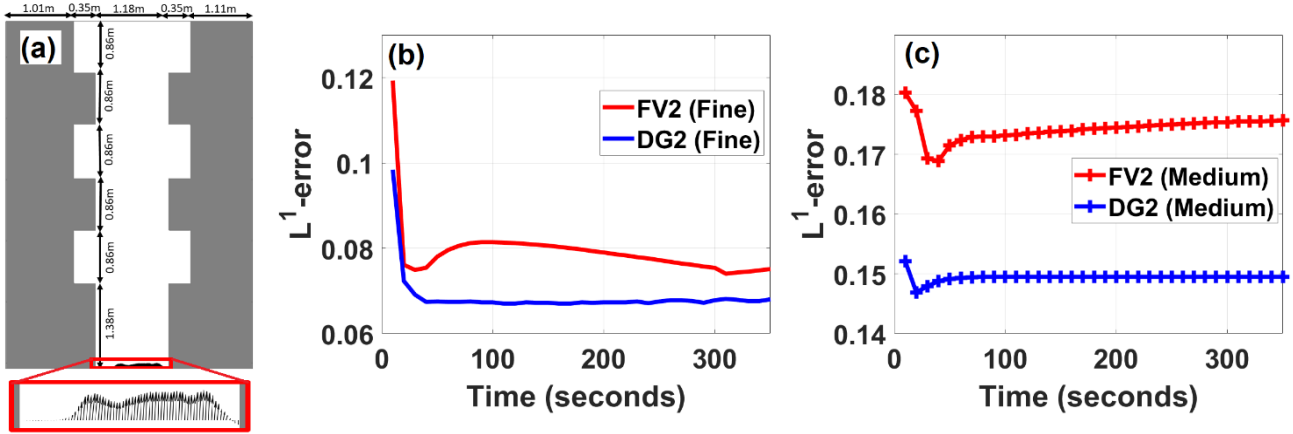

Fig. 5. The left panel (a) shows the test configuration involving the buildings (grey area) and side cavities, where the black arrows in the red box indicate the inflow velocities extracted from the PIV-measured velocity fields along the southern boundary of the flume. The other panels present the time history of $L^1$-norm errors based on the velocity fields produced by FV2 (red line) and DG2 (blue line) up to $t$ = 350 s: (b) fine resolution and (c) medium resolution.

The simulated 2D velocity fields were extracted every 10 s and contrasted against the measured velocity fields, via the $L^1$-norm error, during a 350 s simulation. Fig. 5b-5c shows the time series of the errors generated by the FV2 and DG2 solvers at the fine and medium resolutions, respectively. With both resolutions, the DG2 solver leads to lower error than those with the FV2 solver that stabilise earlier in time. This suggests that the DG2 solver converges faster to predictions that are closer to the measured velocity fields and is less affected by error dissipation as compared to the FV2 solver. At the fine resolution (Fig. 5b), the error sharply decreases with both solvers in the first 20 s, but that of the DG2 solver stabilises by 40 s while that of the FV2 solver requires at least 330 s to become less stable. At the medium resolution (Fig. 5c), the error is relatively higher with both solvers, confirming that resolution coarsening induces a higher error dissipation. Still, the rate of error dissipation with the DG2 solver is lower, stabilises by 60 s, and shows no visually detectable increase. In contrast, the error of the FV2 solver shows a slightly increasing trend suggesting that it is more significantly impacted by the growth of error dissipation. This confirms that the DG2 solver has a better predictive capability than the FV2 solver with resolution coarsening.

The runtime cost for the DG2 solver, at the fine and the medium resolutions, respectively, was quantified at the time instants when the error stagnated (i.e., after 40 s and 60 s) to be 13 and 4.75 minutes, whereas that of the FV2 solver was quantified at 330 s for both resolutions to be 20.1 and 0.27 minutes, respectively. Even at the fine resolution, the DG2 solver is more efficient, because it closely converges to the measured fields earlier (Fig. 5b). At the medium resolution, the DG2 solver is 2.7 times faster but with double the error magnitude (Fig. 5b vs. Fig. 5c), and the 73 times speed-up for the FV2 solver may be signalling too much flattening of small-scale velocity information within its predictions.

The time-averaged 2D spatial velocities were produced and analysed for the predictions made by the DG2 and FV2 solvers during [60 s, 80 s] and [330 s, 350 s], respectively, to be compared with the time-averaged measured velocity fields. Fig. 6 displays the spatial vector map and its corresponding streamlines extracted from the measured velocity fields (Fig. 6c), and those extracted from the time-averaged FV2 and DG2 spatial velocity fields at the fine resolution (Fig. 6a-b) and at the medium resolution (Fig. 6d-e).

In the main channel, both DG2 and FV2 streamlines show consistently linear patterns like the PIV streamlines in the high velocity regions. However, discrepancies in the velocity predictions can be detected in the mainstream flow region surrounding the upstream left cavity where the measured velocities are low with recirculation eddies from the interactions between the inflow and flow reflection from the sharp edges of the upstream left cavity. This leads to complex 3D flows with high turbulence that may be difficult to capture with 2D-SWE solvers in this region. At the fine resolution, the FV2 solver (Fig. 6a) better predicts the recirculation flow length and width but overly elongates the eddies and shifts their position. In contrast, the DG2 solver better predicts the location of the eddies but with narrower length and width of the recirculation flow (Fig. 6b). Such discrepancies in the predictions (Fig. 6a-c) may be expected: the FV2 solver has larger error dissipation (Fig. 5b), leading to elongated recirculation length and, arguably, playing the role of physical dissipation introduced by a turbulence model. With the DG2 solver, the rate of error dissipation is relatively small, not able to replace the dissipation a turbulence model would introduce at the fine resolution. At the medium resolution (Fig. 6d-e), both solvers exhibit higher error dissipation, providing a better capture of the length and width of the recirculation flow, which remains larger with the FV2 solver due to its higher rate of error dissipation (Fig. 5c).

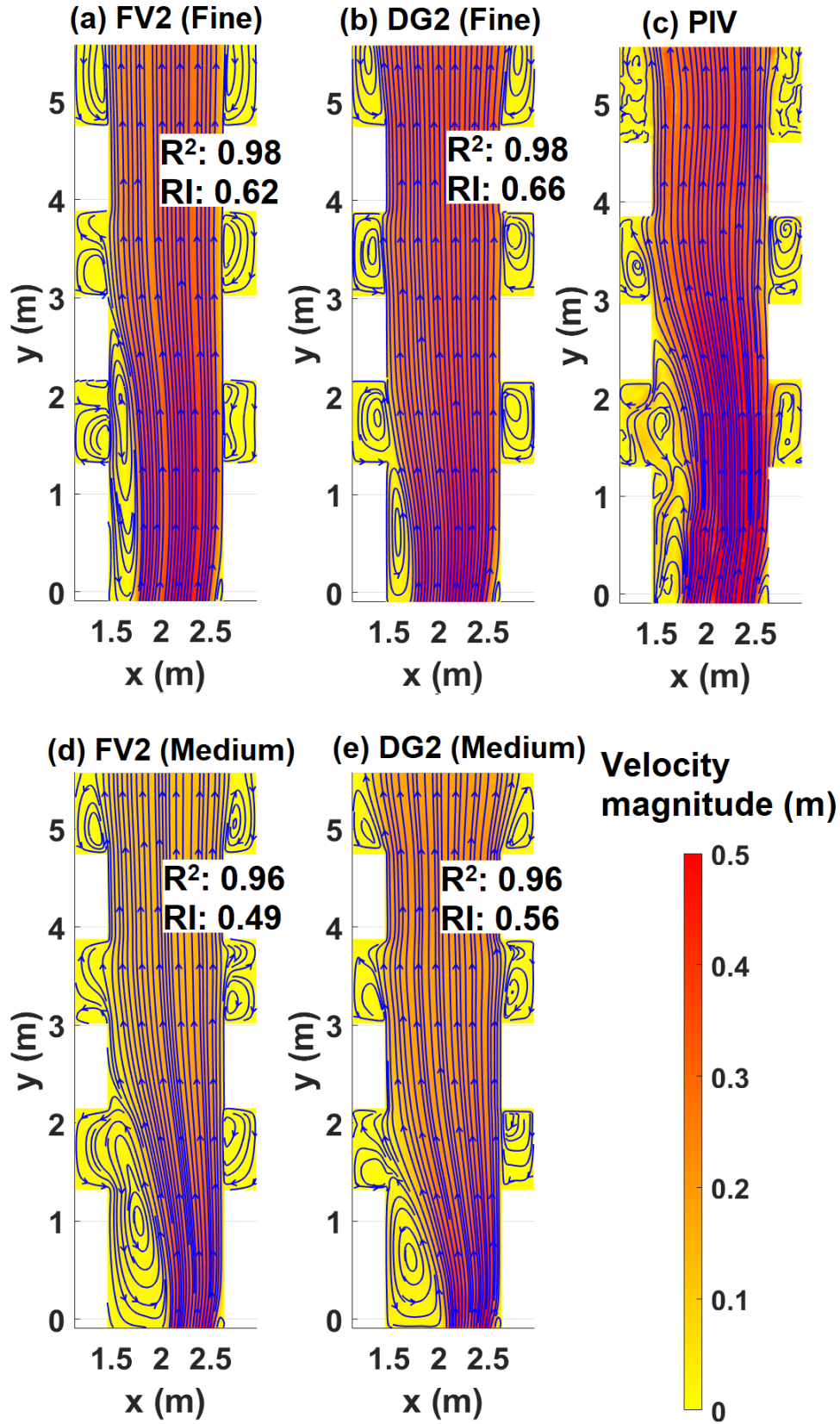

Fig. 6. Spatial vectors and streamlines of 2D time-averaged velocity fields obtained from: (a) the FV2 solver prediction at the fine resolution; (b) the DG2 solver prediction at the fine resolution; (c) the PIV data; (d) the FV2 solver prediction at the medium resolution; and (e) the DG2 solver prediction at the medium resolution.

In the other regions, away from the vicinity of the upstream left cavity, the flow is mostly 2D and less impacted by high turbulence effects. At the fine resolution, the DG2 solver is a clear winner in replicating the eddies within the recirculation flow regions, namely near the southern edge of the left middle cavity and near the northern edge of the right middle cavity. Both FV2 and DG2 solvers lead to similar $R^2$ coefficients, of 0.98, suggesting similar patterns as the measured fields. Despite the DG2 solver underprediction to the mainstream recirculation flow, it led to an RI coefficient, of 0.66, that is higher than that of the FV2 solver, of 0.62, suggesting that the former yields better directional alignment with measured fields around the cavities. At the medium resolution, the DG2 solver outperforms again in a better capture of flow directionality, leading to an RI of 0.56, compared to the 0.49 value from the FV2 solver predictions. These results are quite aligned with those of the previous one, suggesting that the DG2 solver can produce sufficiently accurate velocity fields in the regions where the validity of the depth-integrated 2D-SWE assumption prevails.

### *Flow diversion at a T-junction*

This test case involves subcritical mainstream flow diverting into the lateral branch of a right-angled channel junction. It examines further the predictive performance of DG2 solver, compared to the FV2 solver, in producing spatial velocity fields to measured recirculation width at the entrance of the lateral branch (curved line in Fig. 8a-d) and measured velocity components along a cross-section of flow separation (vertical dashed line in Fig. 8) (Bazin 2013; Bazin et al. 2017; Mignot et al. 2013). Two configurations are considered that are shown in Fig. 7a: one without a 5 cm square obstacle located in the main branch, representing a classical flow diversion featured by a dead zone of recirculation flow in the lateral branch, and another with the obstacle, leading to complex flow features behind the obstacle that impact the flow ahead and past the diversion.

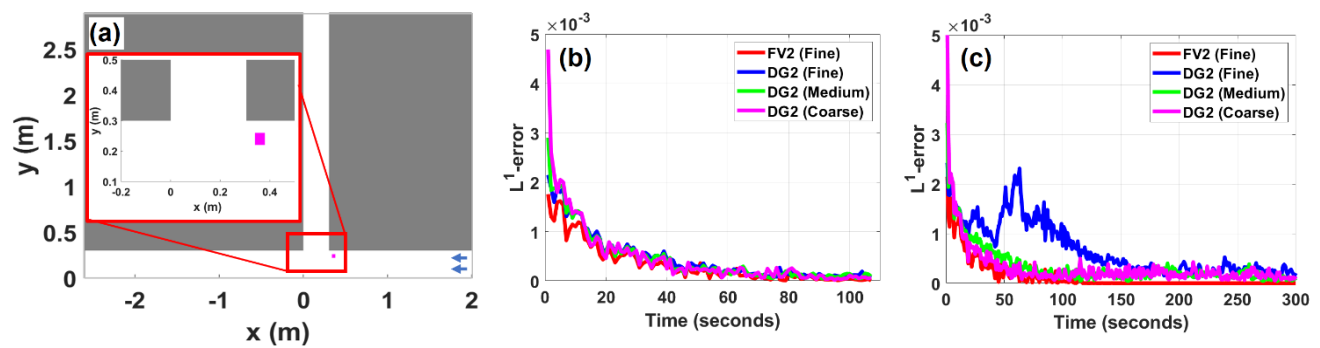

Fig. 7. Panel (a) shows the T-junction spatial domain including the main (white horizontal band) and the lateral (white vertical band) branches and the upstream inflow (arrows) from the east of the main branch. The insert plot (red box) contains the zoomed-in view of the small obstacle position (magenta square). Panels (b) and (c) show the convergence history for the solvers at three different resolutions for the configuration with and without the obstacles, respectively

For both configurations, a steady inflow discharge of 0.002 $m^3s^{-1}$ is considered, associated with a Froude number of 0.23, entering the main branch from the eastern boundary. Following Bazin (2013), fixed water depths of 45.1 mm and 44.6 mm are imposed at the downstream end of the main and the lateral branches, respectively. The bed surface is flat with a Manning's value of $n_M = 0.01$ s $m^{-1/3}$ (Bazin 2013). The simulations, using the DG2 and FV2 solvers, were run on at a resolution of 0.005 m (fine), which was used by alternative 2D-SWE solvers (Bazin et al. 2017). Simulations with the DG2 solver also considered two- and four-times coarser resolutions of 0.01 (medium) and 0.02 m (coarse), respectively. The simulations per configuration were run for a long-enough time until a state of quasi-steadiness was reached. This state was identified by the time history of the difference between the water depth maps across two subsequent time steps, using the $L^1$-norm error, and was assumed after the time instant when the error magnitude and variability are significantly low (i.e., below $10^{-4}$).

Fig. 7b-c includes the time series of the error for the solvers, showing a decrease over time as expected. For the no-obstacle configuration (Fig. 7b), the error trend is somewhat similar with all the solvers, suggesting that the quasi-steady state was reached by 90 s before the simulation stopped at 107 s. To complete the 107 s simulation, the DG2 solver needed around 1.5, 0.3 and 0.1 minutes at the fine, medium and coarse resolutions, respectively, and the FV2 solver at the fine resolution took around 1.25 minutes. Therefore, the velocity fields were time-averaged from the output velocity predictions during [90 s, 100s]. For the obstacle configuration (Fig. 7c), a simulation time of 300 s was required for the DG2 solver to reach quasi-steady state, by 290 s, but the FV2 solver reached it earlier by 100 s. This may be signalling that the predictions of the DG2 solver exhibit more velocity fluctuations. To complete the 300 s simulation, the DG2 solver needed around 2, 0.34 and 0.12 minutes at the fine, medium and coarse resolutions, respectively, and the FV2 solver at the fine resolution was a bit faster than the DG2 solver, around 1.3 minutes. As quasi-steady state was reached by 290 s for all the solvers, the velocity fields were time-averaged from the output velocity predictions during [290 s, 300s].

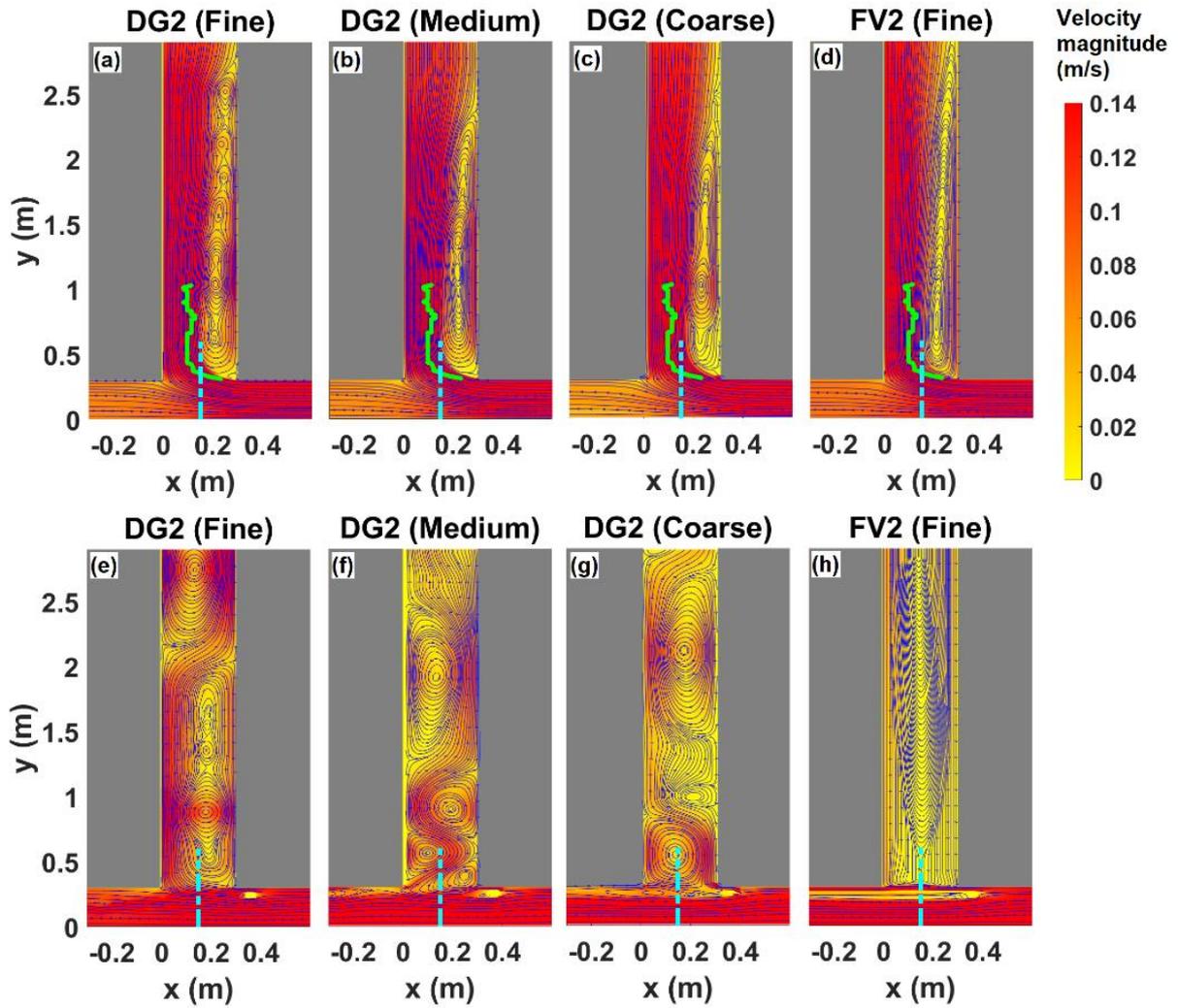

Fig. 8. Streamlines of the time-averaged velocity fields simulated by DG2 solver at the fine, medium and coarse resolution and the FV2 solver at the fine resolution. Panels (a)-(d) show the results of the classical flow separation (no-obstacle); the curved line indicates the measured recirculation width and the dashed line indicate the cross-section ($x = 0.15$ m) of the measured velocity components. Panels (e)-(h) show the results with the obstacle configuration.

Fig. 8a-d displays the streamlines of the time-averaged velocity fields for the classical flow separation (no obstacle). The FV2 solver and the DG2 solver, despite resolution coarsening, predict similar streamlines for the mainstream channel flow prior to, during, and after the flow separation, as well as in the lateral branch away from the dead zone of recirculation flow. In the latter region, all the solvers somewhat predict a similar recirculation length, but the recirculation width simulated at the fine resolution is more aligned with the measured width (Fig. 8a-d). Nonetheless, the DG2 solver could better trail the uneven recirculation width seen in the measured recirculation width (Bazin et al. 2017). This suggests that the flow separation phenomenon is more reliably predicted by the DG2 solver at the fine resolution. Moreover, the DG2 solver at the fine resolution

exhibits a sequence of small-scale eddies that tend to merge and reduce in number with resolution coarsening and were not detected by the FV2 solver. In contrast, the latter predicts a single eddy that is elliptical, elongating towards the downstream end. The superiority of the DG2 solver predictions at the fine resolution can also be seen by comparing its velocity components to the measured ones at the flow separation cross-section of (Fig. 9a-b). The white portion in Fig. 9a-9b spans the part inside the main branch. There, all the solvers fairly predict a uniform spatial distribution of the *u* component that is best captured at the fine resolution, and the rise in the *v* component that is best trailed with the DG2 solver at the fine resolution. Inside the lateral branch, the grey portion in Fig. 9a, the slowing down in the *u* component is best trailed by the DG2 and FV2 solvers at the fine resolution, both capturing the steep rise observed in the measurements. As the flow approaches the dead recirculation zone, all the solvers predict fairly well the *u* component but bigger discrepancies appear in the prediction of the *v* component inside the lateral branch (Fig. 9b). In this region, the steep drop in the *v* velocity is only appropriately predicted by the DG2 solver at the fine resolution. The FV2 solver at the fine resolution, although it predicts a close fit the measurements, smears out the steep drop in the *v* velocity. These analyses suggest that the DG2 and FV2 solvers provide resolved enough velocity field predictions at the fine resolution to describe classical flow diversion through a junction ($R^2$ coefficient around 0.9) but the DG2 solver can provide more informative prediction of the flow phenomena within the separation region.

In Fig. 8e-h, the streamlines of the time-averaged velocity fields are shown for configuration with the obstacle, where all the solvers predict relatively similar flow patterns in the main branch where the flow is not affected by the obstacle. Past the obstacle and along the separation area in the main branch, major discrepancies occur amongst the solver predictions. The predictions of the DG2 solver at the fine resolution are well-resolved to include an eddy formation past the obstacle and a unidirectional flow downstream behind the flow separation area. At the medium resolution, the DG2 predicted eddy past the obstacle is about the same location but with a bit larger length and width, and another eddy is formed downstream away from the flow separation area. At the coarse resolution, the location of the eddy past the obstacle is shifted in the main branch, but with an elongated recirculation zone. With the FV2 solver prediction there is one elongated eddy past the obstacle, stretching towards the downstream with the largest recirculation length and width. The discrepancies in the eddy predictions are expected to occur because the FV2 predicted velocity past the obstacle has a much slower recovery rate compared to the DG2 predicted velocity, causing a longer recirculation length induced by excessive error dissipation (Sun et al. 2022). At the coarse resolution, the DG2 solver may be subject to larger error dissipation than necessary as it predicts a mainstream flow that is quite different than at the medium and fine resolution. At the fine resolution, the DG2 solver predicts streamline patterns along the separation area that are relatively aligned with those observed for the classical flow separation case, suggesting a smaller error dissipation than necessary to replace the presence of a turbulence model. At the medium resolution, the DG2 predictions capture the complex flow interaction along the flow separation area with more details, suggesting that this might be the right resolution to approximate turbulence effects.

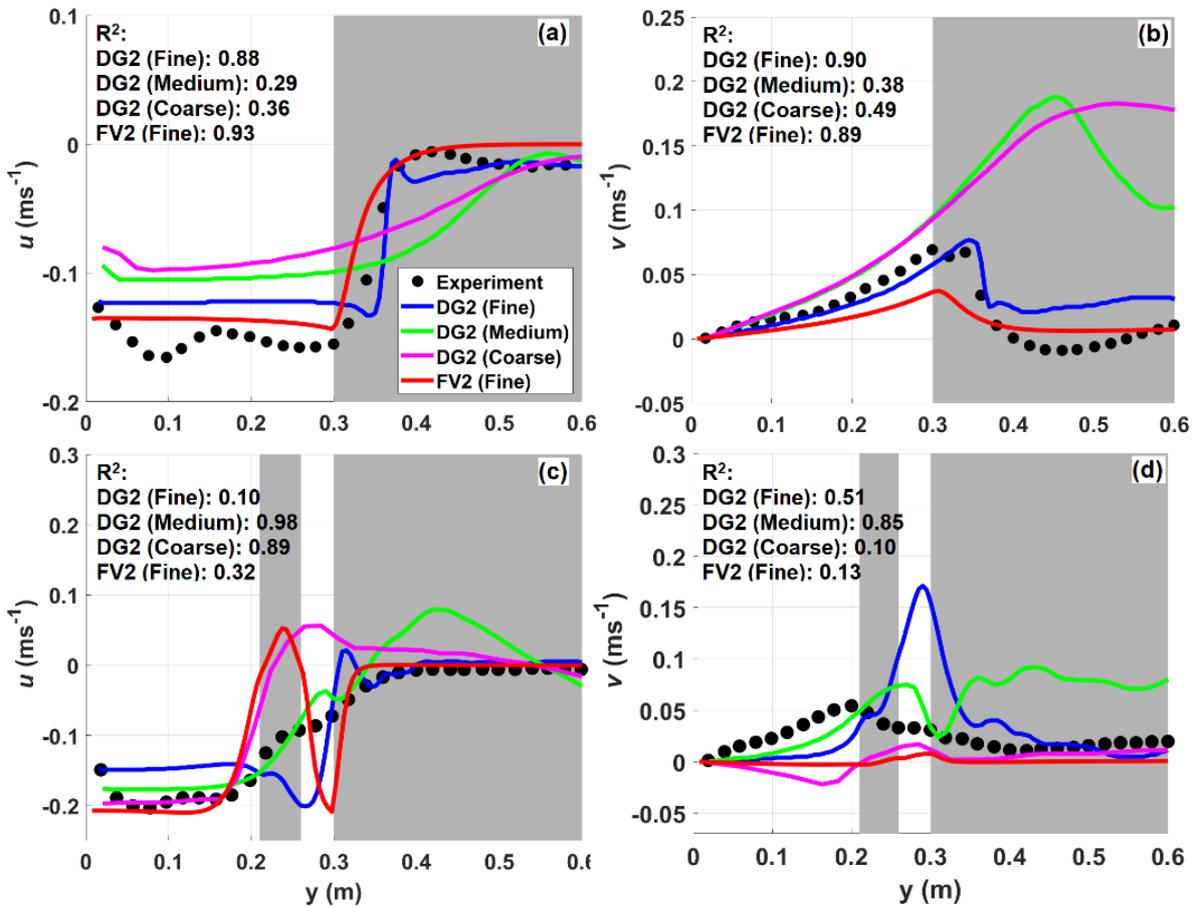

Fig. 9. Lateral profiles of the predicted and measured velocity components at the cross-section $x = 0.15$ m. Panel (a) and (b) show the $u$ and $v$ components, respectively, for the classical flow separation (no-obstacle), while panels (c) and (d) show those associated with the obstacle configuration.

This observation can be further analysed according to the measurements for the predicted $u$ and $v$ velocity components at the cross-section of flow separation. In the main channel including the obstacle (white zone with the grey column in Fig. 9c-9d), the $u$ component is predicted to be uniform for all the solvers except at the region past the obstacle. There, the DG2 solver at the coarse resolution and the FV2 solver suggest the presence of a dead zone caused by its prediction to an elongated eddy leading to slow velocity recovery rate. At the fine resolution, the DG2 predicted at a faster recovery rate for the *u* component than the one seen in the measurements, which are better trailed at the medium resolution. The better performance for the DG2 solver at the medium resolution is clear in the prediction of a *v* component that more closely trails the measurements (Fig. 9d). The FV2 solver predicts no flow entering the lateral branch impacted by the elongated eddy (a zero for the *v* component), whereas the DG2 solver predicts a spurious flow direction at the coarse resolution and exhibits a slight lag in the flow separation at the fine resolution.

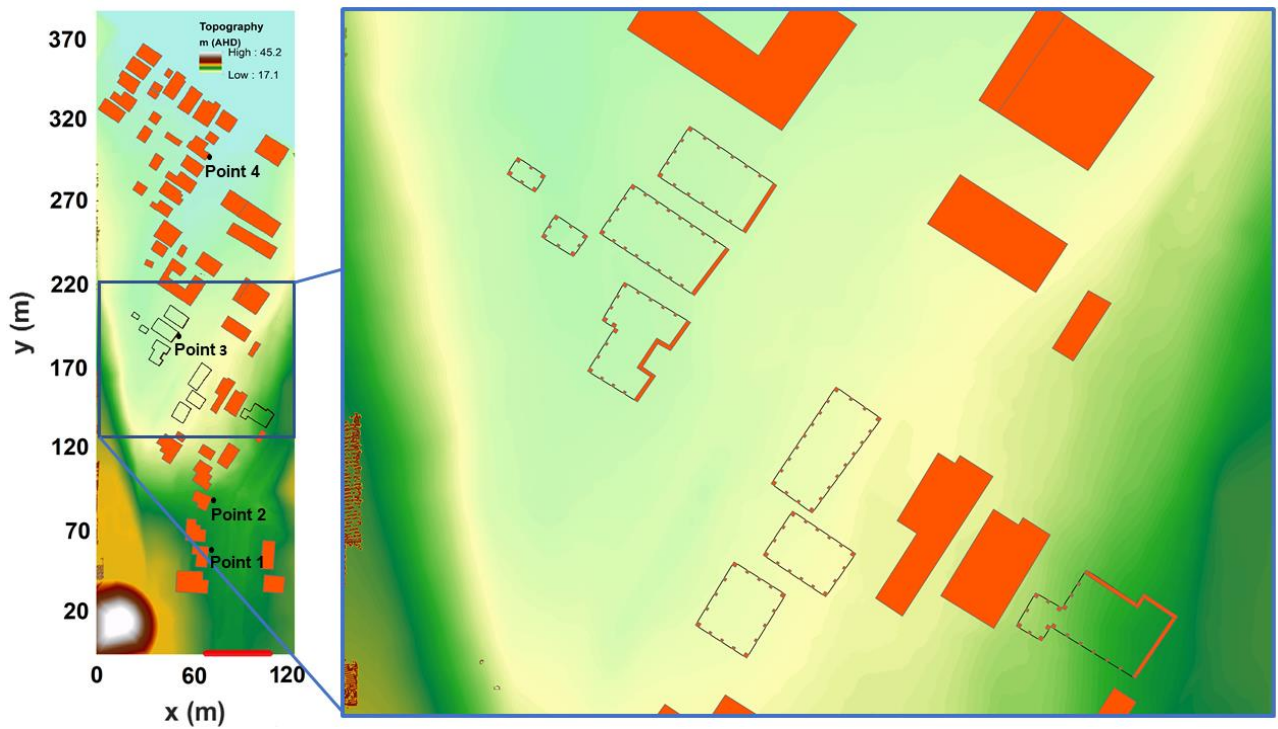

Fig. 10. The left panel illustrates the created digital elevation model for the Morgan-Selwyn floodway prototype, including the buildings on piers (black boxes), water level sampling points (black dots) and the inflow boundary (red line, bottom of the panel). The zoomed-in portion in the right panel shows a view of the walls and the piers within the piered-buildings within which maximum velocity vectors were measured.

## *Flooding in an urban residential area*

This test case particularly investigates the ability of the DG2 and FV2 solvers in reproducing measured maximum velocity vectors within and around buildings with small piers as a result of flash flooding over an urban residential area. The scaled physical model (30:1 horizontal, 9:1 vertical) is 12.5 m long and 5 m wide prototype, replicating the Morgan-Selwyn floodway in Merewether, Australia to recreate the flooding scenario arising from the 'Pasha Bulker' storm (Smith et al. 2016). The inflow is a steady discharge of 19.7 m$^3$/s, which is the estimated peak flood discharge from the storm, flowing through a 38 m opening from the southern boundary (thick line along $y = 0$, Fig. 10, left panel). The inflow propagates over the initially dry topography that is constructed using road base materials. Free outflow conditions are imposed at the downstream north boundary and wall boundary conditions are specified elsewhere.

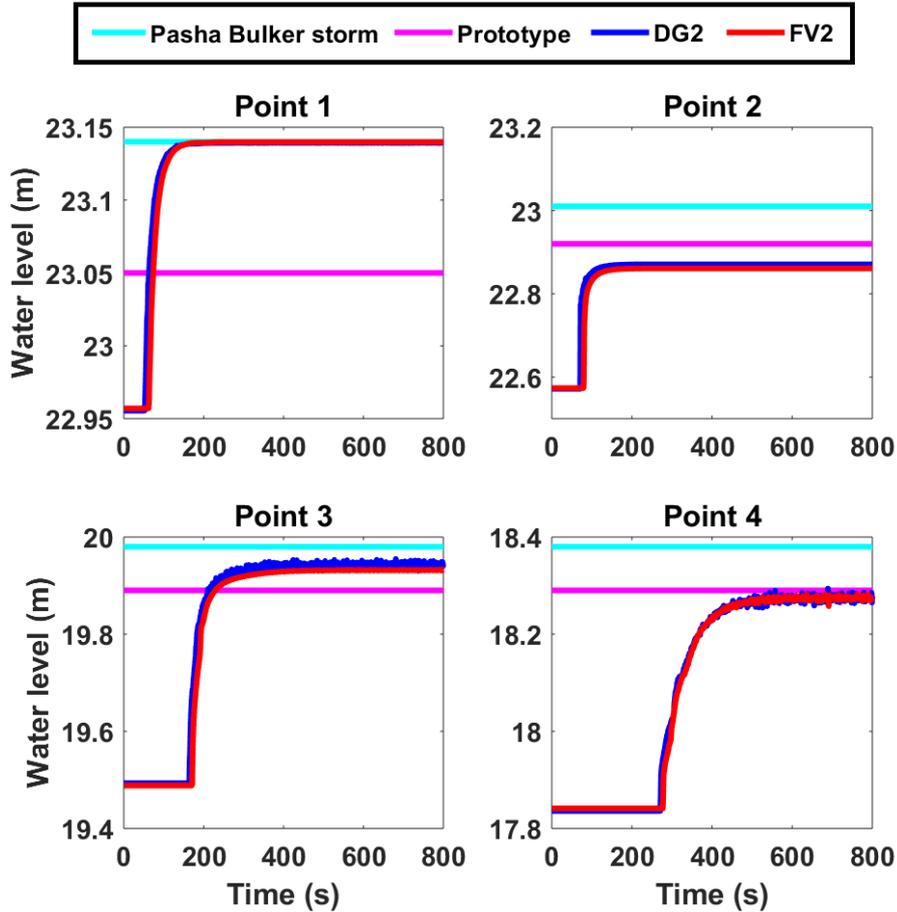

Fig. 11. Time series of the water levels predicted by the DG2 and the FV2 solvers at the four sampling points up to $t = 800$ s, alongside the historical flood level records of the Pasha Bulker storm and the measured water levels from the floodway prototype (Smith et al. 2016).

Although a digital elevation model is available at a fine resolution of 1 cm, a coarser model was created to run the simulations at a resolution of 17.5 cm that is shown in Fig. 10 (Ayog and Sharifian 2022). In the coarser model, the square blocks, and the corner and side building piers were manually added (assumed unsubmerged). The lengths for the corner and side piers are 52 and 35 cm, respectively, which were measured and scaled using detailed photographs (Smith et al. 2016). As in Smith et al. (2016), the Manning's coefficient value was calibrated, to $n_M = 0.042$ s m$^{-1/3}$, to reproduce the historical and measured water level records at sampling point 4 located downstream of the piered buildings (Fig. 10, left panel). The flood simulations were run (www.seamlesswave.com/Merewether) up to 800 s, and a quasi-steady state was reached by 600 s when the water level stagnated at sampling point 4. The DG2 and FV2 solvers took 1.7 minutes and 1.4 minutes to complete the 800 s simulation, respectively. Fig. 11 contains the DG2 and FV2 predicted water level time series at the four sampling points, showing a good agreement with the measured water levels.

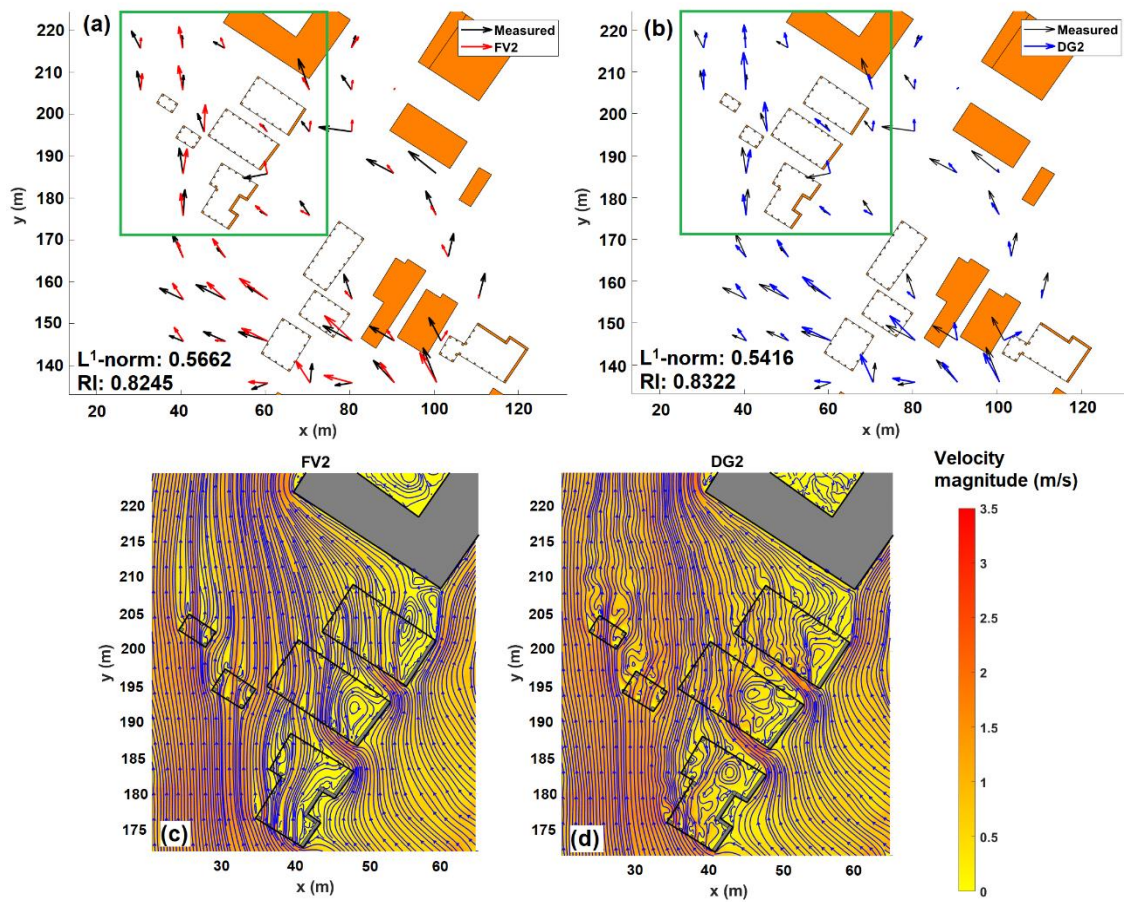

Fig. 12. Panels (a) and (b) show the maximum spatial vectors (zoomed-in portion in Fig. 10) predicted by the FV2 and DG2 solvers, respectively. Panels (c) and (d) show the time-averages velocity streamlines predicted by the FV2 and DG2 solver, respectively, for the area inside the green boxes shown in panels (a) and (b).

To validate the predicted velocity fields around the piered buildings, the simulated maximum velocity vectors produced by the solvers during the 800 s were extracted at a 10 m spacing, consistent with the spacing of the measured maximum velocity vectors. Fig. 12a-b includes a comparison between the velocity vectors extracted from the solver predictions and the measured velocity vectors. Both solvers are seen to fairly capture the measured maximum velocity fields in the region where the piered buildings are located. Quantitatively, the DG2 solver predicts slightly higher RI and lower $L^1$-norm error, indicating that its estimations are closer to and better aligned with the measured maximum velocity fields. Moreover, the predicted velocity fields were time-averaged during [750 s, 800 s], after quasi-steady state was reached, to provide an analysis of the streamlines in the northwest area where the flood wave interaction with the piers is complex (framed area, Fig. 12a-b). As shown in Fig. 12c-d, the DG2 predicted streamlines have more irregularities, exhibit larger swirling patterns, more eddies around and inside the piers. This suggests the DG2 solver could also be a better option for flood simulations, aimed to capture the details of velocity field distribution around topographic structures at sub-meter resolution.

**Conclusions**

This paper demonstrated that second-order discontinuous Galerkin (DG2) solutions to the 2D depth-integrated shallow water equations (2D-SWE) is a better alternative to finite volume solutions to more accurately reproduce quasi-steady spatial velocity distributions at sub-meter resolution. This capability was demonstrated for four selected test cases by comparing the DG2 predicted velocities against those predicted by FV2 or FV3 solvers, with validation against experimentally measured velocities. The test cases involved recirculation eddies past a conical island from vortex shedding, in building side cavities, across a junction structure with/without complex main flow obstruction, and inside and behind small building piers. Quantitative and qualitative analyses of the predicted velocities show that the DG2 solver can at least better reproduce small-scale recirculation eddies in smooth flow regions. These were not detected by the FV2 solver, due to its relatively higher error dissipation, and distorted by the FV3 solver. In the non-smooth regions of complex flow-structure interaction, the DG2 solver can provide a good approximation at a certain range of the resolution where error dissipation is high enough to replace the presence of a turbulence model. Going too fine in resolution leads to well-resolved velocities that include the small-scale eddies in the smooth flow regions, but may mislead the predictions in the complex flow regions where the addition of a turbulence model seems necessary. Sensibly coarsening the resolution could provide a trade-off between the approximation of turbulence effects in the complex flow regions and the preservation of predictions that are sufficiently informative on the flow process in the smooth regions. However, this heuristic approach may work for modelling applications dominated by smooth flow regions that are not significantly impacted by the phenomena from a complex flow region. Within this scope, the DG2 solver is found to be a better choice than finite volume solvers to efficiently acquire more informative velocity fields.

**Data Availability Statement**

Some or all data, models, or code generated or used during the study are available in a repository or online in accordance with funder data retention policies. The code to run the solvers is openly available on LISFLOOD-FP 8.0 (Shaw et al. 2021) and using the local repository of the University of Sheffield (2021). The data to set-up and run the four test cases are made available on the Zenodo (Ayog and Sharifian 2022).

**Acknowledgements**

Georges Kesserwani acknowledges the support of the UK Engineering and Physical Sciences Research Council, grant ID: EP/R007349/1; and Janice Lynn Ayog acknowledges the support from the Malaysian Ministry of Education and Universiti Malaysia Sabah, Malaysia. We thank Xitong Sun for his valuable insights on the analysis of the simulation results.

**References**

Aizinger, V., and Dawson, C. 2002. "A discontinuous Galerkin method for two-dimensional flow and transport in shallow water." *Adv. Water Resour.*, 25, 67-84.

Alcrudo, F., and García-Navarro, P. 1993. "A high-resolution Godunov-type scheme in finite volumes for the 2D shallow-water equations." *Int. J. Numer. Methods Fluids.*, 16, 489-505.

Alvarez-Vázquez, L. J., Martínez, A., Vázquez-Méndez, M. E., and Vilar, M. A. 2008. "An optimal shape problem related to the realistic design of river fishways." *Ecol. Eng.*, 32(4), 293–300. https://doi.org/10.1016/j.ecoleng.2007.10.008.

Aureli, F., Dazzi, S., Maranzoni, A., Mignosa, P., and Vacondio, R. 2015. "Experimental and numerical evaluation of the force due to the impact of a dam-break wave on a structure." *Adv. Water Resour.*, 76, 29–42. https://doi.org/10.1016/j.advwatres.2014.11.009.

Ayog, J. L., and Sharifian, M. K. 2022. "LISFLOOD-DG2 simulation set-up files and additional datasets." Zenodo. https://doi.org/10.5281/zenodo.6914888.

Ayog, J. L., Kesserwani, G., Shaw, J., Sharifian, M. K., and Bau, D. 2021. "Second-order discontinuous Galerkin flood model: Comparison with industry-standard finite volume models." *J. Hydrol.*, 594, 125924. https://doi.org/10.1016/j.jhydrol.2020.125924.

Bassi, F., and Rebay, S. 1997. "High-Order Accurate Discontinuous Finite Element Solution of the 2D Euler Equations." *J. Comput. Phys.*, 138, 251-285.

Bazin, P.-H. 2013. "Flows during floods in urban areas: influence of the detailed topography and exchanges with the sewer system." *PhD thesis.*, :Université Claude Bernard - Lyon I.

Bazin, P.-H., Mignot, E., and Paquier, A. 2017. "Computing flooding of crossroads with obstacles using a 2D numerical model." *J. Hydraul. Res.*, 55(1), 72–84. https://doi.org/ 10.1080/00221686.2016.1217947.

Beisiegel, N., Vater, S., Behrens, J., and Dias, F. 2020. "An adaptive discontinuous Galerkin method for the simulation of hurricane storm surge." *Ocean Dyn.*, 70(5), 641–666. https://doi.org/10.1007/s10236-020-01352-w.

Bernetti, R., Titarev, V., and Toro, E. 2008. "Exact solution of the Riemann problem for the shallow water equations with discontinuous bottom geometry." *J. Comput. Phys.,* 227(6), 3212-3243. https://doi.org/10.1016/j.jcp.2007.11.033.

Bladé, E., Cea, L., Corestein, G., Escolano, E., Puertas, J., Vázquez-Cendón, E., Dolz, J., and Coll, A. 2014. "Iber: herramienta de simulación numérica del flujo en ríos." *Rev. Int. Métodos Numéricos para Cálculo y Diseño en Ing.*, 30(1), 1–10. https://doi.org/10.1016/j.rimni.2012.07.004.

BMT-WBM. 2018. *TUFLOW Classic/HPC User Manual Build 2018-03-AD*.

Bonetti, S., Manoli, G., Manes, C., Porporato, A., & Katul, G. G. (2017). Manning's formula and Strickler's scaling explained by a co-spectral budget model. *Journal of Fluid Mechanics*, 812, 1189–1212. https://doi.org/10.1017/jfm.2016.863

Clare, M., Percival, J., Angeloudis, A., Cotter, C., and Piggott., M. 2021. "Hydro-morphodynamics 2D modelling using a discontinuous Galerkin discretisation." *Comput. Geosci.*, 146, 104658. https://doi.org/10.1016/j.cageo.2020.104658.

Cea, L., Puertas, J., and Vázquez-Cendón, M. E. 2007. "Depth averaged modelling of turbulent shallow water flow with wet-dry fronts." *Arch. Comput. Methods Eng.*, 14(3), 303–341. https://doi.org/10.1007/s11831-007-9009-3.

Chen, S., Garambois, P. A., Finaud-Guyot, P., Dellinger, G., Mosé, R., Terfous, A., and Ghenaim, A. 2018. "Variance based sensitivity analysis of 1D and 2D hydraulic models: An experimental urban flood case." *Environ. Model. Softw.*, 109, 167–181. https://doi.org/10.1016/j.envsoft.2018.08.008.

Cockburn, B., and Shu, C. 2001. "Runge-Kutta Discontinuous Galerkin Methods for Convection Dominated Problems." *J. Sci. Comput.*, 16(3), 173–261. https://doi.org/10.1023/A:1012873910884.

Collecutt, G., and Syme, B. 2017. "Experimental benchmarking of mesh size and time-step convergence for a 1st and 2nd order SWE finite volume scheme." *Proc. 37th IAHR World Congr.*

Dinehart, R. L., and Burau, J. R. 2005. "Repeated surveys by acoustic Doppler current profiler for flow and sediment dynamics in a tidal river." *J. Hydrol.*, 314(1–4), 1-21. https://doi.org/10.1016/j.jhydrol.2005.03.019.

Duan, J. G. 2005. "Two-dimensional model simulation of flow field around bridge piers." *World Water Congr. 2005 Impacts Glob. Clim. Chang. - Proc. 2005 World Water Environ. Resour. Congr.*, Anchorage, Alaska. https://doi.org/10.1061/40792(173)449.

Duran A., and Marche, F. 2014. "Recent advances on the discontinuous Galerkin method for shallow water equations with topography source terms." *Comput. Fluids*, 101, 88–104. Elsevier Ltd. https://doi.org/10.1016/j.compfluid.2014.05.031.

Galland, J. C., Goutal, N., and Hervouet, J.M. 1991. TELEMAC: A new numerical model for solving shallow water equations. *Adv. Wat. Resour.*, 14 (3), 138–148, https://doi.org/10.1016/0309-1708(91)90006-A.

García-Navarro, P., Murillo, J., Fernández-Pato, J., Echeverribar, I., and Morales-Hernández, M. 2019. "The shallow water equations and their application to realistic cases." *Environ. Fluid Mech.*, 19 (5), 1235–1252. https://doi.org/10.1007/s10652-018-09657-7.

Ginting, B. M., and Ginting, H. 2019. "Hybrid Artificial Viscosity–Central-Upwind Scheme for Recirculating Turbulent Shallow Water Flows." *J. Hydraul. Eng.*, 145(12), 04019041.

Gioia, G., and Bombardelli, F. A. (2002). "Scaling and Similarity in Rough Channel Flows." *Phys. Rev. Lett.*, 88(1), 4. https://doi.org/10.1103/PhysRevLett.88.014501

Godunov, S. K. 1959. "A difference method for numerical calculation of discontinuous solutions of the equations of hydrodynamics." *Mat. Sb. (N.S.)*, 47 (89) (3): 271–306.

Guinot, V. 2003. *Godunov-type Schemes: An Introduction for Engineers.* Elsevier, Amsterdam.

Horritt, M. S., Bates, P. D., and Mattinson, M. J. 2006. "Effects of mesh resolution and topographic representation in 2D finite volume models of shallow water fluvial flow." *J. Hydrol.*, 329(1–2), 306–314. https://doi.org/10.1016/j.jhydrol.2006.02.016.

Hou, J., Liang, Q., Zhang, H., and Hinkelmann, R. 2015. "An efficient unstructured MUSCL scheme for solving the 2D shallow water equations." *Environ. Model. Softw.*, 66, 131–152. https://doi.org/10.1016/j.envsoft.2014.12.007.

INRAE. 2021. "2D modelling: Rubar20 and Rubar20TS." *River Hydraul.* Accessed April 5, 2021. https://riverhydraulics.inrae.fr/en/tools/numerical-modelling/2d-modelling-rubar20-and-rubar20ts/.

Jackson, T. R., Apte, S. V., Haggerty, R., and Budwig, R. 2015. "Flow structure and mean residence times of lateral cavities in open channel flows: influence of bed roughness and shape." *Environ. Fluid Mech.*, 15, 1069–1100. https://doi.org/10.1007/s10652-015-9407-2.

JACOBS. 2022. Flood Modeller, 2D ADI Solver. www.floodmodeller.com/2dsolvers. Accessed October 30 2022.

Jodeau, M., Hauet, A., Paquier, A., Le Coz, J., and Dramais, G. 2008. "Application and evaluation of LS-PIV technique for the monitoring of river surface velocities in high flow conditions." *Flow Meas. Instrum.*, 19(2), 117–127. https://doi.org/10.1016/j.flowmeasinst.2007.11.004.

Juez, C., Thalmann, M., Schleiss, A. J., and Franca, M. J. 2018. "Morphological resilience to flow fluctuations of fine sediment deposits in bank lateral cavities." *Adv. Water Resour.*, 115, 44-59. https://doi.org/10.1016/j.advwatres.2018.03.004.

Kärnä, T., Kramer, S. C., Mitchell, L., Ham, D. A., Piggott, M. D., and Baptista, A. M. 2018. "Thetis coastal ocean model: discontinuous Galerkin discretization for the three-dimensional hydrostatic equations." *Geosci. Model Dev.*, 11, 4359–4382. https://doi.org/10.5194/gmd-2017-292.

Kesserwani, G., and Liang, Q. 2012. "Locally limited and fully conserved RKDG2 shallow water solutions with wetting and drying." *J. Sci. Comput.*, 50, 120-144. https://doi.org/10.1007/s10915-011-9476-4

Kesserwani, G. 2013. "Topography discretization techniques for Godunov-type shallow water numerical models: a comparative study." *J. Hydraul. Res.*, 51(4), 351-367. https://doi.org/10.1080/00221686.2013.796574.

Kesserwani, G., and Wang, Y. 2014. "Discontinuous Galerkin flood model formulation: Luxury or necessity?" *Water Resour. Res.*, 50, 6522-6541. https://doi.org/10.1002/2013WR014906.

Kesserwani, G., Ayog, J. L., and Bau, D. 2018. "Discontinuous Galerkin formulation for 2D hydrodynamic modelling: Trade-offs between theoretical complexity and practical convenience." *Comput. Methods Appl. Mech. Eng.*, 342, 710–741. https://doi.org/10.1016/j.cma.2018.08.003.

Kesserwani, G., and Sharifian, M. K. 2020. "(Multi)wavelets increase both accuracy and efficiency of standard Godunov-type hydrodynamic models: Robust 2D approaches." *Adv. Water Resour.*, 144, 103693. https://doi.org/10.1016/j.advwatres.2020.103693.

Kim, D.-H., Lynett, P. J., and Socolofsky, S. A. 2009. "A depth-integrated model for weakly dispersive, turbulent, and rotational fluid flows.", *Ocean Model.*, 27(3-4), 198-214. https://doi.org/10.1016/j.ocemod.2009.01.005.

Kimura, I., and Hosoda, T. 1997. "Fundamental Properties of Flows in Open Channels with Dead Zone.", *J. Hydraul. Eng.*, 123 (2), 98-107. https://doi.org/10.1061/(ASCE)0733-9429(1997)123:2(98).

Krivodonova, L., Xin, J., Remacle, J. F., Chevaugeon, N., and Flaherty, J. E. 2004. "Shock detection and limiting with discontinuous Galerkin methods for hyperbolic conservation laws." *Appl. Numer. Math.*, 48(3–4), 323–338. https://doi.org/10.1016/j.apnum.2003.11.002.

Krivodonova, L. 2007. "Limiters for high-order discontinuous Galerkin methods.", *J. Comput. Phys.*, 226, 879-896. https://doi.org/10.1016/j.jcp.2007.05.011.

Kubatko, E. J., Westerink, J. J., and Dawson, C. 2006. "hp Discontinuous Galerkin methods for advection dominated problems in shallow water flow." *Comput. Methods Appl. Mech. Eng.*, 196(1–3), 437–451. https://doi.org/10.1016/j.cma.2006.05.002.

Lambrechts, J., Humphrey, C., McKinna, L., Gourge, O., Fabricius, K. E., Mehta, A. J., Lewis, S., and Wolanski., E. 2010. "Importance of wave-induced bed liquefaction in the fine sediment budget of Cleveland Bay, Great Barrier Reef." *Estuar. Coast. Shelf Sci.*, 89 (2), 154–162. Elsevier Ltd. https://doi.org/10.1016/j.ecss.2010.06.009.

LaRocque, L. A., Elkholy, M., Hanif Chaudhry, M., and Imran, J. 2013. "Experiments on Urban Flooding Caused by a Levee Breach." *J. Hydraul. Eng.*, 139(9), 960–973. https://doi.org/10.1061/(asce)hy.1943-7900.0000754.

Le Bars, Y., Vallaeys, V., Deleersnijder, É., Hanert, E., Carrere, L., and Channelière, C. 2016. "Unstructured-mesh modeling of the Congo river-to-sea continuum." *Ocean Dyn.*, 66 (4), 589–603. https://doi.org/10.1007/s10236-016-0939-x.

Le, H.-A., Lambrechts, J., Ortleb, S., Gratiot, N., Deleersnijder, E., and Soares-Frazão, S. 2020. "An implicit wetting–drying algorithm for the discontinuous Galerkin method: application to the Tonle Sap, Mekong River Basin." *Environ. Fluid Mech.*, 20, 923-951. https://doi.org/10.1007/s10652-019-09732-7.

Legleiter, C. J., and Kinzel, P. J. 2020. "Inferring surface flow velocities in sediment-laden alaskan rivers from optical image sequences acquired from a helicopter." *Remote Sens.*, 12(8), 1–28. https://doi.org/10.3390/RS12081282.

Liang, Q., and Marche, F. 2009. "Numerical resolution of well-balanced shallow water equations with complex source terms." *Adv. Water Resour.*, 32 (6), 873–884. https://doi.org/10.1016/j.advwatres.2009.02.010.

Lloyd, P. M., and Stansby, P. K. 1997. "Shallow-Water Flow around Model Conical Islands of Small Side Slope. II: Submerged." *J. Hydraul. Eng.*, 123(12), 1068–1077.

Lynett, P. J., Gately, K., Wilson, R., Montoya, L., Arcas, D., Aytore, B., Bai, Y., Bricker, J. D., Castro, M. J., Cheung, K. F., David, C. G., Dogan, G. G., Escalante, C., González-Vida, J. M., Grilli, S. T., Heitmann, T. W., Horrillo, J., Kânoğlu, U., Kian, R., Kirby, J. T., Li, W., Macías, J., Nicolsky, D. J., Ortega, S., Pampell-Manis, A., Park, Y. S., Roeber, V., Sharghivand, N., Shelby, M., Shi, F., Tehranirad, B., Tolkova, E., Thio, H. K., Velioğlu, D., Yalçıner, A. C., Yamazaki, Y., Zaytsev, A., and Zhang, Y. J. 2017. "Inter-model analysis of tsunami-induced coastal currents." *Ocean Model.*, 114, 14–32.

Macías, J., Castro, M. J., and Escalante, C. 2020. "Performance assessment of the Tsunami-HySEA model for NTHMP tsunami currents benchmarking. Laboratory data." *Coast. Eng.*, :Elsevier B.V., 158(July 2018), 103667.

Mazur, R., Kałuża, T., Chmist, J., Walczak, N., Laks, I., and Strzeliński, P. 2016. "Influence of deposition of fine plant debris in river floodplain shrubs on flood flow conditions – The Warta River case study." *Phys. Chem. Earth.*, 94, 106–113. https://doi.org/10.1016/j.pce.2015.12.002.

Mignot, E., and Brevis, W. 2020. "Coherent Turbulent Structures within Open-Channel Lateral Cavities." *J. Hydraul. Eng.*, 146(2), 04019066. https://doi.org/10.1061/(asce)hy.1943-7900.0001698.

Mignot, E., Zheng, C., Gaston, D., Chi, W.-L., Riviere, N., and Bazin, P.-H. 2013. "Impact of topographic obstacles on the discharge distribution in open-channel bifurcations." *J. Hydrol.*, 10–19. https://dx.doi.org/10.1016/j.jhydrol.2013.04.023.

Momplot, A., Lipeme Kouyi, G., Mignot, E., Rivière, N., and Bertrand-Krajewski, J. L. 2017. "Typology of the flow structures in dividing open channel flows." *J. Hydraul. Res.*, 55(1), 63–71. https://doi.org/10.1080/00221686.2016.1212409.

Mulamba, T., Bacopoulos, P., Kubatko, E. J., and Pinto, G. F. 2019. "Sea-level rise impacts on longitudinal salinity for a low-gradient estuarine system." *Clim. Change*, 152, 533–550. https://doi.org/10.1007/s10584-019-02369-x.

Navas-Montilla, A., Juez, C., Franca, M. J., and Murillo, J. 2019. "Depth-averaged unsteady RANS simulation of resonant shallow flows in lateral cavities using augmented WENO-ADER schemes." *J. Comput. Phys.*, 395, 511-536. https://doi.org/10.1016/j.jcp.2019.06.037.

NTHMP (National Tsunami Hazard Mitigation Program). 2016. *Report on the 2015 NTHMP Current*

*Modeling Workshop. NTHMP Publ. Resour.*, Portland, Oregon.

Ouro, P., Juez, C., and Franca, M. 2020. "Drivers for mass and momentum exchange between the main channel and river bank lateral cavities.", *Adv. Water Resour.*, 137, 103511. https://doi.org/10.1016/j.advwatres.2020.103511.

Özgen-Xian, I., Xia, X., Liang, Q., Hinkelmann, R., Liang, D., and Hou, J. 2021. "Innovations towards the next generation of shallow flow models." *Adv. Water Resour.*, 149(January), 103867. https://doi.org/10.1016/j.advwatres.2021.103867.

Pagliara, S., and Carnacina, I. 2013. "Bridge pier flow field in the presence of debris accumulation." *Proc. Inst. Civ. Eng. Water Manag.*, 166(4), 187–198. https://doi.org/10.1680/wama.11.00060.

Pham Van, C., de Brye, B., Deleersnijder, E., Hoitink, A. J. F., Sassi, M., Spinewine, B., Hidayat, H., and Soares-Frazão, S. 2016. "Simulations of the flow in the Mahakam river–lake–delta system, Indonesia." *Environ. Fluid Mech.*, 16 (3), 603–633. https://doi.org/10.1007/s10652-016-9445-4.

Rubinato, M. 2015. "Physical scale modelling of urban flood systems." *PhD thesis.*, :The University of Sheffield.

Rubinato, M., Shucksmith, J., Martins, R., and Kesserwani, G. 2021. "Particle Image Velocimetry (PIV) dataset for the parking lot configuration with a closed manhole." Zenodo. https://doi.org/10.5281/zenodo.4596731.

Ruiz-Villanueva, V., Wyżga, B., Mikuś, P., Hajdukiewicz, M., and Stoffel, M. 2017. "Large wood clogging during floods in a gravel-bed river: the Długopole bridge in the Czarny Dunajec River, Poland." *Earth Surf. Process. Landforms.*, 42, 516–530. https://doi.org/10.1002/esp.4091.

Schaal, K., Bauer, A., Chandrashekar, P., Pakmor, R., Klingenberg, C., and Springel, V. 2015. "Astrophysical hydrodynamics with a high-order discontinuous Galerkin scheme and adaptive mesh refinement." *Mon. Notices Royal Astron. Soc.*, 453(4), 4278–4300. https://doi.org/10.1093/mnras/stv1859.

Shaw, J., Kesserwani, G., Neal, J., Bates, P., and Sharifian, M. K. 2021. "LISFLOOD-FP 8.0 : the new discontinuous Galerkin shallow water solver for multi-core CPUs and GPUs." *Geosci. Model Dev. Discuss.* https://doi.org/10.5194/gmd-14-3577-2021.

Shettar, A. S., and Murthy, K. K. 1996. "A numerical study of division of flow in open channels." *J. Hydraul. Res.*, 34(5), 651–675. https://doi.org/10.1080/00221689609498464.

Smith, G. P., Rahman, P. F., and Wasko, C. 2016. "A comprehensive urban floodplain dataset for model benchmarking." *Int. J. River Basin Manag.*, 14(3), 345–356. https://doi.org/10.1080/15715124.2016.1193510.

Sweby, P. K. 1984. "High Resolution Schemes Using Flux Limiters for Hyperbolic Conservation Laws." *SIAM J. Numer. Anal.*, 21(5), 995–1011.

Sun, X., Kesserwani, G., Sharifian, M.K., and Stovin, V. 2022. Simulation of laminar to transitional wakes past cylinders with discontinuous Galerkin shallow water solutions, *J. Hydraul. Res.*, in press.

Syme, W. J. 2008. "Flooding in Urban Areas - 2D Modelling Approaches for Buildings and Fences." *9th Natl. Conf. Hydraul. Water Eng. Hydraul.*, 25–32. Barton, A.C.T.: Engineers Australia.

Toro, E. F. 1989. "A weighted average flux method for hyperbolic conservation laws.", *Proc. R. Soc. Bond. A,* 423, 401-418.

Toro, E. F. 2001. *Shock-Capturing Methods for Free-Surface Shallow Flows*. West Sussex, UK: John Wiley & Sons.

Toro, E. F., and García-Navarro, P. 2007. "Godunov-type methods for free-surface shallow flows: A review." *J. Hydraul. Res.*, 45(6), 736–751. https://doi.org/10.1080/00221686.2007.9521812.

University of Sheffield. 2021. "LISFLOOD-FP8.0 with DG2 and GPU solvers." *SEAMLESS-WAVE*. Accessed April 5, 2021. https://www.seamlesswave.com/LISFLOOD8.0.

UASCE. 2021. HEC-RAS 2D User's Manual. Welcome to HEC-RAS. Version 6.3. Accessed October 30, 2022. https://www.hec.usace.army.mil/confluence/rasdocs/r2dum/latest.

van den Abeele, K., Broeckhoven, T., and Lacor, C. 2007. "Dispersion and dissipation properties of the 1D spectral volume method and application to a p-multigrid algorithm." *J. Comput. Phys.*, 224 (2), 616–636. https://doi.org/10.1016/j.jcp.2006.10.022.

van Leer, B. 2006. "Upwind high-resolution methods for compressible flow: from donor cell to residual-distribution." *Commun. Comput. Phys.*, 1(2), 192-206.

Wang, J. W., and Liu, R. X. 2005. "Combined finite volume-finite element method for shallow water equations." *Comput. Fluids*, 34 (10), 1199–1222. https://doi.org/10.1016/j.compfluid.2004.09.008.

Wang, Z. J., Fidkowski, K., Abgrall, R., Bassi, F., Caraeni, D., Cary, A., Deconinck, H., Hartmann, R., Hillewaert, K., Huynh, H. T., Kroll, N., May, G., Persson, P.-O., Leer, B. van, and Visbal, M. 2013. "High-order CFD methods: current status and perspective." *Int. J. Numer. Methods Fluids*., 72, 811–845. https://doi.org/10.1002/fld.3767.

Wintermeyer, N., Winters, A. R., Gassner, G. J., and Warburton, T. 2018. "An entropy stable discontinuous Galerkin method for the shallow water equations on curvilinear meshes with wet/dry fronts accelerated by GPUs.", *J. Comput. Phys.*, 375, 447-480. https://doi.org/10.1016/j.jcp.2018.08.038.

Winters, A. R., and Gassner, G. J. 2015. "A comparison of two entropy stable discontinuous Galerkin spectral element approximations for the shallow water equations with non-constant topography." *J. Comput. Phys.*, 301, 357-376.

Wood, D., Kubatko, E. J., Rahimi, M., Shafieezadeh, A., and Conroy, C. J. 2020. "Implementation and evaluation of coupled discontinuous Galerkin methods for simulating overtopping of flood defenses by storm waves." *Adv. Water Resour.*, 136, 103501. https://doi.org/10.1016/j.advwatres.2019.103501.

Wu, X., Kubatko, E. J., and Chan, J. 2021. "High-order entropy stable discontinuous Galerkin methods for the shallow water equations: Curved triangular meshes and GPU acceleration." *Comput. Math. with Appl.*, 82, 179-199. https://doi.org/10.1016/j.camwa.2020.11.006.

Zhang, M. P., Shu, C.-W. 2005. "An analysis of and a comparison between the discontinuous Galerkin and the spectral finite volume methods." *Comput. Fluids*, 34(2005), 581-592. https://doi.org/10.1016/j.compfluid.2003.05.006.

Zhang, Y. J., Priest, G., Allan, J., and Stimely, L. 2016. "Benchmarking an Unstructured-Grid Model for Tsunami Current Modeling." *Pure Appl. Geophys.*, 173(12), 4075–4087.

Zhou, T., Li, Y., and Shu, C.-W. 2001. "Numerical Comparison of WENO Finite Volume and Runge-Kutta Discontinuous Galerkin Methods." *J. Sci. Comput.,* 16(2), 145-171.